\documentstyle[aps,12pt]{revtex}

\begin{document}
\author{J.R. Morris\thanks{%
E-mail address: jmorris@iunhaw1.iun.indiana.edu}}
\address{{\it Department of Chemistry/Physics/Astronomy, }\\
{\it Indiana University Northwest, 3400 Broadway,}\\
{\it \ Gary, Indiana 46408}}
\title{{\bf Excitation Fields in a Superconducting Global String}}
\maketitle

\begin{abstract}
A model of a straight superconducting global cosmic string is examined in a
setting wherein the string supports a charge/current pulse described by a
travelling wave along the string. Linearized field equations are obtained
for fluctuations of the scalar and vector fields of the theory, and a set of
approximate particular solutions are found for the case in which the linear
charge density and the current of the string have equal magnitudes. Although
the equations of motion seem to suggest that the scalar and vector
excitation fields are massive inside the string core, the particular
solutions show that they behave as effectively massless fields which
propagate at the speed of light along the string along with the primary
charge/current pulse. The effect of the mass parameter is to modulate the
radial profile of the excitation fields. The vector excitation field
generates radial and angular components for both the electric and magnetic
fields, but the particular solutions do not describe the emission or
absorption of electromagnetic radiation from the string.

\newpage\
\end{abstract}

\section{Introduction}

A superconducting global string can arise from a situation wherein, by the
Witten mechanism\cite{Witten}, the spontaneous symmetry breaking of a global
$U(1)$ symmetry is accommodated by the formation of a scalar condensate
which breaks a gauged $U(1)$ symmetry in the core of the string. The string
can thus be endowed with a charge per unit length $Q$ and a current $I$
which are determined by the complex scalar field which forms the condensate
and the vector gauge field which is coupled to it. $Q$ and $I$ may be either
time independent or time-dependent, and in particular $Q$ and $I$ can be
described in terms of travelling waves along the string\cite{Dayi}. A basic
ansatz\cite{Peter} for a straight superconducting global string can be
constructed where the moduli of the scalar fields and the vector gauge field
structure function depend only upon the radial distance $r$ from the center
of the string and the phase of the scalar condensate field depends upon the
position $z$ along the string, along with the time $t$.

Here, attention is focused upon a situation where relatively small
fluctuations of the fields about their basic ansatz values are considered.
These small fluctuations, or excitation fields, are treated as classical
fields, and linearized equations for the excitation fields are obtained. By
implementing an ansatz, some approximate particular solutions of the linear
excitation field equations can be extracted. In particular, approximate
solutions describing the excitation fields are found for the specific case
where a charge/current pulse travels along the string with $Q^2-I^2=0$ .
Although the equations of motion seem to imply that the scalar and vector
excitation fields are massive inside the string core, the particular
solutions that are found show that they are effectively massless fields that
propagate at the speed of light along the string with the charge/current
pulse. The excitation fields give rise to additional charge and current
densities, along with electromagnetic fields. The electric and magnetic
fields are each found to possess both a radial and an angular component, but
there is no absorption or emission of electromagnetic radiation from the
string.

The model describing the superconducting global string is presented in
section II. The excitation field equations are described in section III, and
approximate solutions are obtained in section IV. A summary and discussion
forms section V.

\section{The Superconducting Global String}

\subsection{General Field Equations}

Consider a straight, infinite global cosmic string\cite{Vilenkin}$,$\cite
{VilenkinEverett} lying along the $z$ axis (we use cylindrical coordinates $%
(r,\theta ,z)$), with a complex-valued scalar string field $\chi $
interacting with a complex-valued scalar field $\phi $ coupled to a $U(1)$
gauge field $A_\mu $, as described by the Lagrangian
\begin{equation}
L=\partial ^\mu \chi ^{*}\partial _\mu \chi +(D^\mu \phi )^{*}(D_\mu \phi
)-V-\frac 14F^{\mu \nu }F_{\mu \nu }\text{ ,}  \label{one}
\end{equation}

\noindent where
\begin{equation}  \label{two}
D_\mu =\nabla _\mu +ieA_\mu \,,\,\,\,\,\,\,\,\,\,\,F_{\mu \nu }=\partial
_\mu A_\nu -\partial _\nu A_\mu \,,
\end{equation}

\noindent with the potential $V$ given by
\begin{equation}  \label{three}
V=\lambda \left( \chi ^{*}\chi -\frac 12\eta ^2\right) ^2+2f\left( \chi
^{*}\chi -\frac 12\eta ^2\right) \phi ^{*}\phi +m^2\phi ^{*}\phi +g\left(
\phi ^{*}\phi \right) ^2,
\end{equation}

\noindent where $\nabla _\mu $ represents the ordinary spacetime covariant
derivative, and a metric with signature $(+,-,-,-)$ is used. The coupling
constants $\lambda $, $f$, and $g$, as well as the $\phi $ particle mass $m$%
, are assumed to be positive, real quantities. The stable vacuum state of
the theory is located by $\left| \chi \right| =\eta $, $\phi =0$, and the $%
\chi $ particle mass is $m_\chi =(2\lambda )^{\frac 12}\,\,\eta $.

The general field equations which follow from (\ref{one}) are given by
\begin{equation}  \label{four}
\Box \chi +\left[ 2\lambda \left( \chi ^{*}\chi -\frac 12\eta ^2\right)
+2f\phi ^{*}\phi \right] \chi =0\,\,,
\end{equation}
\begin{equation}  \label{five}
\begin{array}{ll}
\Box \phi & -e^2\phi A^\mu A_\mu +ie\left( 2A^\mu \partial _\mu \phi +\phi
\nabla _\mu A^\mu \right) \\
& +\left[ 2f\left( \chi ^{*}\chi -\frac 12\eta ^2\right) +m^2+2g\phi
^{*}\phi \right] \phi =0,
\end{array}
\end{equation}
\begin{equation}  \label{six}
\begin{array}{lll}
\nabla _\mu F^{\mu \nu } & = & \Box A^\nu =eJ_T^\nu \\
& = & \,ie\left[ \phi ^{*}\partial ^\nu \phi -\phi \partial ^\nu \phi
^{*}\right] \,-2e^2\phi ^{*}\phi A^\nu \,,
\end{array}
\end{equation}

\noindent where $\Box \equiv \nabla _\mu \nabla ^\mu $ and we have chosen
the Lorentz gauge $\nabla _\mu A^\mu =0$, and the conserved total current
density is
\begin{equation}  \label{seven}
\begin{array}{lll}
J_T^\nu & = & i\left[ \phi ^{*}(D^\nu \phi )-\phi (D^\nu \phi )^{*}\right]
\\
& = & i\left[ \phi ^{*}\partial ^\nu \phi -\phi \partial ^\nu \phi
^{*}\right] -2e\phi ^{*}\phi A^\nu \,\,,\,\,\,\,\,\,\,\,\,\,\nabla _\nu
J_T^\nu =0\,\,.
\end{array}
\end{equation}

\subsection{Parametrization and Primary Fields}

Let us consider a parametrization of the various fields as given by
\begin{equation}  \label{eight}
\chi =\frac 1{\sqrt{2}}\left[ R(r)+G(\vec r,t)\right] \,e^{i\theta }\,\,,
\end{equation}
\begin{equation}  \label{nine}
\phi =\frac 1{\sqrt{2}}\left[ F(r)+H(\vec r,t)\right] \,e^{i\psi (t,z)}\,\,,
\end{equation}
\begin{equation}  \label{ten}
A_\mu =B_\mu +a_\mu (\vec r,t),\,\,\,\,\,\,\,\,\,\,B_\mu \equiv \frac 1e%
\left[ P(r)-1\right] \partial _\mu \psi (t,z)\,\,,
\end{equation}
\begin{equation}  \label{eleven}
F_{\mu \nu }=B_{\mu \nu }+f_{\mu \nu }\,,\,\,\,\,\,\,\,B_{\mu \nu }\equiv
\partial _\mu B_\nu -\partial _\nu B_\mu \,,\,\,\,\,\,\,\,f_{\mu \nu }\equiv
\partial _\mu a_\nu -\partial _\nu a_\mu \,\,.\,
\end{equation}

\noindent The total conserved current $J_T^\nu $ can be written as
\begin{equation}  \label{twelve}
J_T^\nu =J^\nu +j^\nu \,\,,
\end{equation}

\noindent where $J^\nu $ is obtained from (\ref{seven}) upon setting $G=0$, $%
H=0$, and $a_\mu =0$, so that in this case one obtains
\begin{equation}  \label{thirteen}
\begin{array}{ll}
\chi =\frac 1{\sqrt{2}}R(r)\,e^{i\theta }\,\,, & \phi =\frac 1{\sqrt{2}}%
F(r)\,e^{i\psi (t,z)}\,\,, \\
A_\mu =B_\mu \,=\frac 1e\left[ P(r)-1\right] \partial _\mu \psi \,,\,\,\, &
J^\nu =-F^2P\partial ^\nu \psi \,\,.
\end{array}
\end{equation}

\noindent The fields in (\ref{thirteen}) will be referred to as the {\it %
primary} fields, and the fields
\begin{equation}  \label{fourteen}
\chi _1\equiv \frac 1{\sqrt{2}}G(\vec r,t)\,e^{i\theta
}\,\,,\,\,\,\,\,\,\,\,\,\,\phi _1\equiv \frac 1{\sqrt{2}}H(\vec r%
,t)\,e^{i\psi (t,z)}\,\,,\,\,\,\,\,\,\,\,\,\,a_\mu =A_\mu -B_\mu
\end{equation}

\noindent will be referred to as the {\it excitation} fields.

A basic ansatz for the superconducting global string can be built from the
primary fields, so that upon setting the excitation fields equal to zero,
the primary field equations which follow from (\ref{four}) -- (\ref{six})
can be written as
\begin{equation}  \label{fifteen}
\frac 1r\partial _r\left( r\partial _rR\right) -\frac R{r^2}-\left[ \lambda
\left( R^2-\eta ^2\right) +fF^2\right] R=0\,\,,
\end{equation}
\begin{equation}  \label{sixteen}
\frac 1r\partial _r\left( r\partial _rF\right) +FP^2\partial _\mu \psi
\partial ^\mu \psi -\left[ f\,\left( R^2-\eta ^2\right) +m^2+gF^2\right]
F=0\,\,,
\end{equation}
\begin{equation}  \label{seventeen}
\frac 1r\partial _r\left( r\partial _rP\right) -e^2F^2P=0\,\,.
\end{equation}

\noindent The potential in (\ref{three}), in terms of the primary scalar
fields, becomes
\begin{equation}  \label{eighteen}
V=\frac 14\lambda \left( R^2-\eta ^2\right) ^2+\frac 12f\,\left( R^2-\eta
^2\right) F^2+\frac 12m^2F^2+\frac 14gF^4\,\,.
\end{equation}

\noindent Inside the string, $R\rightarrow 0$ as $r\rightarrow 0$, so that
in the string core the potential $V$ is minimized by a value of $F$ given by
\begin{equation}  \label{nineteen}
F_0=\left[ \frac{\left( f\eta ^2-m^2\right) }g\right] ^{\frac 12}\,\,,
\end{equation}

\noindent whereas in the true vacuum region, as $r\rightarrow \infty $, and $%
R\rightarrow \eta ,$ the potential $V$ is minimized by a value of $F$ given
by $F=0$. Therefore, we take the boundary conditions for the primary field
equations (\ref{fifteen} -- \ref{seventeen}) describing the superconducting
global string to be
\begin{equation}  \label{twenty}
R\rightarrow 0\,\,,\,\,\,\,\,\,\,F\rightarrow
F_0\,\,,\,\,\,\,\,\,\,\,\,\,P\rightarrow 1\,\,,\,\,\,\,\,\,\,\,\,\,\partial
_rP\rightarrow 0\,\,,\,\,\,\,\,\,\,\,\,\,\text{as }r\rightarrow 0\,\,,
\end{equation}
\begin{equation}  \label{twenty-one}
R\rightarrow \eta \,\,,\,\,\,\,\,\,\,\,\,\,F\rightarrow
0\,\,,\,\,\,\,\,\,\,\,\,\,B_{\mu \nu }\rightarrow 0\,\,,\,\,\,\,\,\,\,\,\,\,%
\text{as }r\rightarrow \infty \,\,.
\end{equation}

\noindent It is furthermore assumed that the quantity $\partial _\mu \psi
\partial ^\mu \psi $ in (\ref{sixteen}) is given by
\begin{equation}  \label{twenty-two}
\partial _\mu \psi \partial ^\mu \psi =K\,\,,
\end{equation}

\noindent where $K$ is a constant, and the radius of the string is taken to
be $r_0\approx m_\chi ^{-1}=\left[ \left( 2\lambda \right) ^{\frac 12}\eta
\right] ^{-1}$.

\subsection{Charge, Current, and Electromagnetic Fields}

For the basic superconducting global string ansatz described by (\ref
{thirteen}) with $j^\nu =0$, the primary current density $J^\nu
=-F^2(r)P(r)\partial ^\nu \psi (t,z)$ is conserved so that
\begin{equation}  \label{twenty-three}
\nabla _\mu J^\mu =0\Rightarrow \left( \partial _0^2-\partial _z^2\right)
\psi (t,z)\equiv \Box _2\psi (t,z)=0\,\,,
\end{equation}

\noindent where $\Box _2\equiv \left( \partial _0^2-\partial _z^2\right) $.
The linear charge density and the current that are enclosed within a radius $%
r$ are
\begin{equation}  \label{twenty-four}
Q_{en}(r)=2\pi e\int_0^rJ^0\,\,r^{\prime }\,dr^{\prime }=-2\pi eJ(r)\partial
_0\psi \,\,,
\end{equation}
\begin{equation}  \label{twenty-five}
I_{en}(r)=2\pi e\int_0^rJ^z\,\,r^{\prime }\,dr^{\prime }=2\pi eJ(r)\partial
_z\psi \,\,,
\end{equation}

\noindent respectively, with $J(r)\equiv \int_0^rF^2P\,\,r^{\prime
}\,dr^{\prime }$. The total (primary) charge per unit length and the total
(primary) current (due to the primary current density $J^\mu $) carried by
the string are $Q=Q_{en}(\infty )\approx Q_{en}(r_0)$ and $I=I_{en}(\infty
)\approx I_{en}(r_0)$, respectively, where $r_0\sim m_\chi ^{-1}=\left(
\sqrt{2\lambda }\,\,\eta \right) ^{-1}$ is the string core radius. The
electric and magnetic fields that are generated by the primary current
density are therefore given by
\begin{equation}  \label{twenty-six}
E_r(r)=\frac{Q_{en}(r)}{2\pi r}\,\,,\,\,\,\,\,\,\,\,\,\,B_\theta (r)=\frac{%
I_{en}(r)}{2\pi r}\,\,.
\end{equation}

By (\ref{twenty-three}) the general solution for $\psi $ can be expressed in
terms of travelling waves,
\begin{equation}  \label{twenty-seven}
\psi (t,z)=\psi _{+}(t+z)+\psi _{-}(t-z)\,\,.
\end{equation}

\noindent In terms of the light cone coordinates $\xi _{\pm }\equiv t\pm z$,
we have $\psi =\psi _{+}(\xi _{+})+\psi _{-}(\xi _{-})$. However, when the
constraint given by (\ref{twenty-two}) is taken into account, we have for
the general solutions $\psi =\omega t+kz+const$ with $\omega ^2-k^2=K$ for
the case $K\neq 0$, whereas for the case $K=0$,
\begin{equation}  \label{twenty-eight}
\psi =\psi (\xi )\,\,,\,\,\,\,\,\,\,\,\,\,\text{where {\it either}}%
\,\,\,\,\,\xi =\xi _{+}\,\,\,\,\text{{\it or}}\,\,\,\,\,\xi =\xi
_{-}\,\,,\,\,\,\,\,(K=0)\,\,.
\end{equation}

\noindent Let us further require that the current density $J_\mu $, and
therefore $\partial _\mu \psi $, be a {\it bounded} function of $\xi $, so
that for the case $K=0$, $\psi $ can include the description of a pulse with
arbitrary shape travelling either up the string in the $+z$ direction or
down the string in the $-z$ direction, in addition to a static charge and
current. (Note that, for $K=0$, $Q_{en},\,\,I_{en},\,\,E_r,\,\,$and $%
B_\theta $ can be considered as functions of $r,\,t,\,$and $z$, in general.)

\section{Excitation Fields}

\subsection{Linearized Excitation Field Equations}

Now consider the case where the excitation fields described by (\ref
{fourteen}) are nonvanishing, but are generally considered to be
sufficiently small in comparison to the primary fields. Using the
parametrized fields of (\ref{eight}) --(\ref{eleven}) the field equations
given by (\ref{four}) -- (\ref{seven}) yield the linearized excitation field
equations
\begin{equation}  \label{twenty-nine}
\Box G+\frac G{r^2}+2i\partial _\mu G\partial ^\mu \theta +\lambda G\left(
3R^2-\eta ^2\right) +fF\left( FG+2RH\right) =0\,\,,
\end{equation}
\begin{equation}  \label{thirty}
\begin{array}{c}
\Box H-KH+i\left( 2\partial _\mu H\,\partial ^\mu \psi +H\Box \psi \right)
-e^2B_\mu \left( B^\mu H+2a^\mu F\right) \\
-2e\left[ B^\mu \left( H\partial _\mu \psi -i\partial _\mu H\right) +a^\mu
\left( F\partial _\mu \psi -i\partial _\mu F\right) \right] \\
+f\left[ \left( R^2-\eta ^2\right) H+2FRG\right] +\left( m^2+3gF^2\right) H=0
\end{array}
\,\,,
\end{equation}
\begin{equation}  \label{thirty-one}
\nabla _\mu f^{\mu \nu }=\Box a^\nu =ej^\nu =-eF\left[ 2H\left( \partial
^\nu \psi +eB^\nu \right) +eFa^\nu \right] \,\,,
\end{equation}

\noindent and $\nabla _\mu j^\mu =0$ implies that
\begin{equation}  \label{thirty-two}
\partial _\nu \left( FH\right) \partial ^\nu \psi +FH\Box \psi +e\nabla _\nu
\left( FHB^\nu \right) +F\partial _\nu F\,a^\nu =0\,\,.
\end{equation}

\subsection{Approximations and Ansatz}

Equations (\ref{twenty-nine}) -- (\ref{thirty-two}) can be simplified
considerably by making some simple assumptions and implementing an ansatz.
Let it first be assumed that solutions exist for which there is a negligible
back reaction upon the global string field, i.e. let us make the
approximation $G\approx 0$. This is seen to be an approximate solution of (%
\ref{twenty-nine}) provided that $fFRH\approx 0$, which is satisfied both
inside the string core ($R\rightarrow 0$) and outside the string ($%
FH\rightarrow 0$). It is also convenient to regard the global string as a
tube of false vacuum with a radius $r_0\approx \left[ \left( 2\lambda
\right) ^{\frac 12}\eta \right] ^{-1}$, and make the approximations
\begin{equation}  \label{thirty-three}
F\approx \left\{
\begin{array}{c}
F_0 \\
0
\end{array}
\right\} ,\,\,\,\,\,\,\,\,R\approx \left\{
\begin{array}{c}
0 \\
\eta
\end{array}
\right\} ,\,\,\,\,\,\,\,\,P\approx \left\{
\begin{array}{c}
1 \\
0
\end{array}
\right\} ,\,\,\,\,\,\,\,\,\,\,\,\,\,\,\,for\,\,\,\,\,\left\{
\begin{array}{c}
r\leq r_0 \\
r>r_0
\end{array}
\right\} .
\end{equation}

\noindent Along with these approximations, let us impose the ansatz
conditions
\begin{equation}  \label{thirty-four}
\partial _\mu \psi \partial ^\mu \psi
=K=0\,\,,\,\,\,\,\,\,\,\,\,\,\,\,\,\,\,a^\mu \partial _\mu
F=0\,\,,\,\,\,\,\,\,\,\,\,\,a^\mu \partial _\mu \psi =0\,\,.
\end{equation}

\noindent The first condition in (\ref{thirty-four}) allows travelling waves
of charge and current to propagate either up the string or down the string,
since $\psi =\psi (\xi )$, where either $\xi =\xi _{+}$ or $\xi =\xi _{-}$,
and the second condition, which is a simplifying condition, implies that $%
a^r=0$ for the general case where $F=F(r)$ or $\partial _\mu F\neq 0$. The
last condition of (\ref{thirty-four}), along with (\ref{thirty-two}) and $%
a^\mu \partial _\mu F=0$, implies, for the general case where $P\neq 0$,
i.e. $eB_\mu \neq -\partial _\mu \psi $, the auxiliary condition
\begin{equation}  \label{thirty-five}
\partial _\mu H\,\partial ^\mu \psi =0\,\,.
\end{equation}

\subsection{Approximate Excitation Field Equations}

Using the fact that $\Box \psi =0$, the approximation $G\approx 0$, along
with (\ref{thirty-three}) -- (\ref{thirty-five}), the linearized equations
reduce greatly. Inside the string core, for $r\leq r_0$, we have
\begin{equation}  \label{thirty-six}
\Box H+M^2H=0\,\,,\,\,\,\,\,\,\,\,\,\,\,\,\,\,\,\,\,\,\,\,M\equiv (2g)^{%
\frac 12}\,F_0\,\,,
\end{equation}
\begin{equation}  \label{thirty-seven}
\nabla _\mu f^{\mu \nu }=\Box a^\nu =ej^\nu =-eF_0\left( 2H\partial ^\nu
\psi +eF_0a^\nu \right) \,\,,
\end{equation}

\noindent where $\nabla _\mu j^\mu \equiv 0$, and outside the string core,
for $r>r_0$,
\begin{equation}  \label{thirty-eight}
\Box H+m^2H=0\,\,,
\end{equation}
\begin{equation}  \label{thirty-nine}
\nabla _\mu f^{\mu \nu }=\Box a^\nu =ej^\nu =0\,\,.
\end{equation}

\noindent In addition, $H$ must satisfy the auxiliary condition $\partial
_\mu H\,\partial ^\mu \psi =0$, given by (\ref{thirty-five}). Note that the
field $H(\vec r,t)$ thus has the appearance of a massive Higgs field, with
mass $M=(2g)^{\frac 12}F_0$ inside the string and mass $m$ outside the
string. However, solutions shall be found for which the field $H$ behaves as
an effectively massless field that can propagate along the string at the
speed of light, with the mass parameters serving to modulate the radial
profile of the field. This type of behavior will be seen to arise in
response to the constraint $\partial _\mu H\,\partial ^\mu \psi =0$ that has
been imposed. Similarly, the photon excitation field $a^\mu $ has the
appearance of a massive vector field inside the string core and a massless
vector field outside the string, but the solutions to be found will suggest
that the vector field $a^\mu $ can actually behave as an effectively
massless field propagating along the string at the speed of light along with
the scalar field $H$.

\section{Solutions}

We regard $H(\vec r,t)$ as being small compared to $F(r)$, i.e. $|H|\ll F(r)$%
. In addition, the constraint $\partial _\mu H\,\partial ^\mu \psi =0$ must
be satisfied. It follows from the boundary conditions given by (\ref{twenty}%
) and (\ref{twenty-one}) that $H$ and $a_\mu $ must satisfy the boundary
conditions
\begin{equation}  \label{forty}
\begin{array}{c}
|H|\ll F_0,\,\,\,\,\,a_\mu \rightarrow 0,\,\,\,\,\,\,\,\, \text{as}%
\,\,\,\,\,r\rightarrow 0,\,\,\,\,\, \\
\text{ }H\rightarrow 0,\,\,\,\,|\,a_\mu |\ll |B_\mu |,\,\,\,\,\,\text{as
\thinspace }r\rightarrow \infty .
\end{array}
\end{equation}

\subsection{Scalar Field}

Let us write $H(\vec r,t)$ in the form $H=H(r,\theta ,\xi )$. Then from $%
\partial _\mu \psi \partial ^\mu \psi =0$ it follows that $\psi =\psi (\xi )$%
, as given by (\ref{twenty-eight}). From the condition given by (\ref
{thirty-five}), $\partial _\mu H\partial ^\mu \psi =0\Rightarrow \partial
_{+}H\partial _{-}\psi +\partial _{-}H\,\partial _{+}\psi =0$, where $%
\partial _{\pm }\equiv \partial /\partial \xi _{\pm }$\thinspace \thinspace
. Therefore, if $\psi =\psi (\xi _{+})$, then $\partial _{-}\psi =0$, which
implies that $\partial _{-}H=0$, and therefore $H=H(\xi _{+})$. On the other
hand, if $\psi =\psi (\xi _{-})$, then $\partial _{+}\psi =0$, so that $%
\partial _{+}H=0$, and therefore $H=H(\xi _{-})$. In either case, if $\psi
=\psi (\xi )$, then $H=H(\xi )$, where either $\xi =\xi _{+}$ or $\xi =\xi
_{-}$, and consequently
\begin{equation}  \label{forty-one}
\Box _2H(r,\theta ,\xi )\equiv (\partial _0^2-\partial _z^2)H(r,\theta ,\xi
)=4\partial _{+}\partial _{-}H(r,\theta ,\xi )=0\,\,.
\end{equation}

\noindent  Therefore, in general, both $\psi $ and $H$ can describe pulses
that propagate along together at the speed of light either up the string or
down the string.

Writing $H_l(r,\theta ,\xi )=h_l(\xi )q_l(r)\sin (l\theta )$, where $%
l=0,1,2,\cdots $, (\ref{thirty-six}) reduces, for $r\leq r_0$, to
\begin{equation}  \label{forty-two}
\partial _r^2q_l+\frac 1r\partial _rq_l-\left( M^2+\frac{l^2}{r^2}\right)
q_l=0\,\,,
\end{equation}

\noindent which is solved by $q_l=\alpha _lI_l(Mr)$, where $I_l$ is a
hyperbolic Bessel function of order $l$, and is zero (for $l\neq 0$) at $r=0$%
, and $\alpha _l$ is a constant. (Note that $H_0(r,\theta ,\xi )=0$.) The
solution $H_l$ can then be displayed as
\begin{equation}  \label{forty-three}
H_l(r,\theta ,\xi )=\alpha _lh_l(\xi )I_l(Mr)\sin (l\theta
)\,\,,\,\,\,\,\,\,\,\,\,\,(\text{\thinspace \thinspace }r\leq r_0)
\end{equation}

\noindent where $h(\xi )$ is a sufficiently bounded, but otherwise
arbitrary, function of $\xi $. The general solution for $H$ inside the
string core is given by a linear superposition, $H(r,\theta ,\xi
)=\sum_lH_l(r,\theta ,\xi )$.

Outside the string core, for $r>r_0$, a similar type of solution exists,
except that $M^2$ must be replaced by $m^2$, and the allowed radial
solutions that conform to the boundary conditions are the hyperbolic Bessel
functions $K_l(mr)$, which are finite, rapidly decreasing functions of $r$
outside the string core. The exterior solutions are therefore given by
\begin{equation}  \label{forty-four}
H_l(r,\theta ,\xi )=\beta _lh_l(\xi )K_l(mr)\sin (l\theta
)\,\,,\,\,\,\,\,\,\,\,\,\,(r>r_0)\,.
\end{equation}

The continuity of $H(r,\theta ,\xi )$ demands that the interior and exterior
solutions be joined at $r=r_0$, and since $I_l(Mr)$ is an increasing
function of $r$, while $K_l(mr)$ is a decreasing function of $r$, we
conclude that, within the context of the approximations employed, the scalar
excitation field $H(r,\theta ,\xi )$ is concentrated at the surface of the
string, i.e. at $r=r_0$, and travels in a nondispersive way along the string
at the speed of light together with the primary charge/current pulse arising
from the current density $J_\mu $. The profile of the scalar field along the
string, given by $h_l(\xi )$, can be a function of arbitrary shape, subject
to the requirement that $H(r,\theta ,\xi )$ be small compared to the primary
field $F(r)$.

\subsection{Vector Field}

In addition to the boundary conditions given by (\ref{forty}), the vector
field $a^\mu (\vec r,t)$ must be further restricted by requiring that the
electric field $\vec e$ and the magnetic field $\vec b$ that are generated
from $a^\mu $ be well defined at $r=0$. In particular, we require
\begin{equation}  \label{forty-five}
e_r\rightarrow 0,\,\,\,\,\,e_\theta \rightarrow 0,\,\,\,\,\,b_r\rightarrow
0,\,\,\,\,\,b_\theta \rightarrow 0,\,\,\,\,\,\,\,\,\,\,\text{as }%
r\rightarrow 0\,\,.
\end{equation}

For the interior region of the string, $r\leq r_0$, equations (\ref
{thirty-four}), (\ref{thirty-five}), and (\ref{thirty-seven}) are seen to be
solved by
\begin{equation}  \label{forty-six}
a_l^\mu (r,\theta ,\xi )=\gamma H_l(r,\theta ,\xi )\partial ^\mu \psi
=\gamma \alpha _lh_l(\xi )I_l(Mr)\sin (l\theta )\partial ^\mu \psi
\,,\,\,\,\,\,(r\leq r_0)
\end{equation}

\noindent where $H_l$ is the interior scalar field solution given by (\ref
{forty-three}), provided that the constant $\gamma $ is given by
\begin{equation}  \label{forty-seven}
\gamma =\frac{2eF_0}{(M^2-e^2F_0^2)}=\frac{2e}{(2g-e^2)F_0}\,\,.
\end{equation}

\noindent Note that for $l=0$, $a_0^\mu =0$. For this solution to satisfy
the conditions given by (\ref{forty-five}) we must require that $\alpha _1=0$%
, which implies that $\beta _1=0$, due to continuity of the scalar field $H$
at $r=r_0$. We therefore have
\begin{equation}  \label{forty-eight}
\alpha _1=\beta _1=0,\,\,H_0(r,\theta ,\xi )=H_1(r,\theta ,\xi
)=0,\,\,a_0^\mu (r,\theta ,\xi )=a_1^\mu (r,\theta ,\xi )=0.
\end{equation}

Outside the string, for $r>r_0$, let $a^\mu $ be displayed as $a_l^\mu
(r,\theta ,\xi )=c_lh_l(\xi )f_l(r)\sin (l\theta )\partial ^\mu \psi $,
where $c_l$ is a constant. Then (\ref{thirty-nine}) reduces to
\begin{equation}  \label{forty-nine}
\partial _r^2f_l+\frac 1r\partial _rf_l-\frac{l^2}{r^2}f_l=0\,\,,
\end{equation}

\noindent which is solved by $f_l(r)=r^{-l}$ ($l\neq 0$). Upon requiring
continuity of the interior and exterior solutions at $r=r_0$, we must have $%
c_0=c_1=0$ so that $a_0^\mu =a_1^\mu =0$ for all $r$. The exterior solution
for the vector field is then given by
\begin{equation}  \label{fifty}
\begin{array}{ll}
a_l^\mu (r,\theta ,\xi ) & =c_lh_l(\xi )f_l(r)\sin (l\theta )\partial ^\mu
\psi \\
& =c_lh_l(\xi )r^{-l}\sin (l\theta )\partial ^\mu \psi
\end{array}
\,\,,\,\,\,\,\,(\,\,r>r_0).
\end{equation}

Since the vector field increases radially for $r\leq r_0$ and then decreases
radially for $r>r_0$, $a^\mu $ is seen to be concentrated at $r=r_0$, with a
tail that falls off as $r^{-l}$ outside the string.

\subsection{Electromagnetic Fields}

The electromagnetic fields exterior to the string can be obtained from (\ref
{fifty}), and are given by
\begin{equation}  \label{fifty-one}
e_{r,l}=-\partial _ra_l^0=lc_lh_l(\xi )r^{-(l+1)}\sin (l\theta )\partial
^0\psi =\frac lra_l^0\,\,,
\end{equation}
\begin{equation}  \label{fifty-two}
e_{\theta ,l}=-\frac 1r\partial _\theta a_l^0=-lc_lh_l(\xi )r^{-(l+1)}\cos
(l\theta )\partial ^0\psi \,=-\frac lr\cot (l\theta )\,a_l^0\,\,,
\end{equation}
\begin{equation}  \label{fifty-three}
b_{r,l}=\frac 1r\partial _\theta \,a_l^z=lc_lh_l(\xi )r^{-(l+1)}\cos
(l\theta )\partial ^z\psi =\frac lr\cot (l\theta )\,a_l^z\,\,,
\end{equation}
\begin{equation}  \label{fifty-four}
b_{\theta ,l}=-\partial _ra_l^z=lc_lh_l(\xi )r^{-(l+1)}\sin (l\theta
)\partial ^z\psi =\frac lra_l^z\,\,.
\end{equation}

The Poynting vector associated with these fields is
\begin{equation}  \label{fifty-five}
\begin{array}{ll}
\vec s_l=\vec e_l\times \vec b_l & =\hat z\,l^2c_l^2h_l^2(\xi
)r^{-2(l+1)}\partial ^0\psi \,\partial ^z\psi =\hat z\, \frac{l^2}{r^2}%
a_l^0a_l^z \\
& =\mp \hat z\,\left( lc_lh_l(\xi )r^{-(l+1)}\right) ^2\left( \partial _{\pm
}\psi \right) ^2\,\,,
\end{array}
\,
\end{equation}

\noindent which indicates that the electromagnetic power flows along the
string with the excitation pulse.

For a superconducting string formed during a Grand Unified Theory (GUT)
symmetry breaking phase transition, with a GUT mass scale characterized by $%
\eta $, the linear charge density $Q$ and the current $I$ can be enormous%
\cite{Witten}, with $I_{\max }\sim O(e\eta )$. Let us consider, for example,
a scalar field phase function $\psi (\xi )$, where we take $\xi =\xi
_{-}=(t-z)$, for definiteness. Then $\partial _0\psi =\partial _{-}\psi
=\partial _\xi \psi ,$ and $\partial _z\psi =-\partial _\xi \psi $, and by (%
\ref{twenty-four}) and (\ref{twenty-five}) the primary linear charge density
and current are given by $Q=I\approx -2\pi eJ(r_0)\partial _\xi \psi $,
where $J(r_0)\sim \frac{F_0^2\,r_0^2}2\sim \frac{F_0^2}{4\lambda \eta ^2}$,
so that $I\sim -\left( \frac{e\pi F_0^2}{2\lambda \eta ^2}\right) \partial
_\xi \psi $, which implies that
\begin{equation}  \label{fifty-six}
\partial _\xi \psi \sim -\frac I{e\pi F_0^2r_0^2}\sim -I\left( \frac{%
2\lambda \eta ^2}{e\pi F_0^2}\right) \,\,.
\end{equation}

\noindent By (\ref{fifty-one}) -- (\ref{fifty-four}) it then follows that
\begin{equation}  \label{fifty-seven}
|\vec e_l|=|\vec b_l|=lr^{-(l+1)}\,|c_lh_l(\xi )\partial _\xi \psi |\sim
\frac l{r^{(l+1)}}\,|c_lh_l(\xi )|\,\left( \frac{2\lambda \eta ^2}{e\pi F_0^2%
}\right) \,I\,\,.
\end{equation}

\noindent As a specific example, consider the phase function $\psi =\omega
\xi +\psi _p(\xi )$, where $\xi =t-z$ and $\psi _p$ is a bounded function of
$\xi $ describing a pulse travelling along the string. By defining the
constant current $I_0\equiv -\omega \,\left( \frac{e\pi F_0^2}{2\lambda \eta
^2}\right) $ and the pulse current $I_p(\xi )\equiv -\left( \frac{e\pi F_0^2%
}{2\lambda \eta ^2}\right) \partial _\xi \psi _p(\xi )$, (\ref{fifty-seven})
gives field strengths $|\vec e_l|=|\vec b_l|\sim lr^{-(l+1)}\left( \frac{%
2\lambda \eta ^2}{e\pi F_0^2}\right) \left| c_lh_l(\xi )\left( I_0+I_p(\xi
)\right) \right| $, with $h_l(\xi )$ being a bounded function of $\xi $ (so
that $H(\vec r,t)$ cannot grow too large in comparison to $F(r)$) satisfying
$\partial _\mu h_l\partial ^\mu \psi =0$. Because the current $I$ can be so
large, the fields $\vec e$ and $\vec b$, although perhaps quite small in
comparison to the primary fields $E_r=\frac Q{2\pi r}$ and $B_\theta =\frac I%
{2\pi r}\,$ , can themselves be quite large. It is furthermore interesting
to note that while the primary electric field $\vec E$ has only a
nonvanishing radial component and the primary magnetic field $\vec B$ has
only a nonvanishing angular component, the fields $\vec e$ and $\vec b$ each
have nonvanishing radial and angular components, in general.

\section{Summary and Discussion}

A model of a superconducting global string has been examined where the
spontaneous breaking of a $U(1)$ global symmetry, giving rise to the global
string, is accompanied by the formation of a charged scalar condensate
(coupled to a $U(1)$ gauge field) in the core of the string. The string
supports a linear charge density $Q$ and carries a current $I$. Attention
has been focused upon the special case for which, by (\ref{twenty-four}) and
(\ref{twenty-five}), $Q^2-I^2=\left( 2\pi eJ(r)\right) ^2\partial _\mu \psi
\partial ^\mu \psi =0$, which is a Lorentz invariant condition. This type of
condition is invoked, for example, for certain chiral vorton models\cite
{Davis},\cite{Morris}, and has the feature that, by (\ref{twenty-six}), the
electric and magnetic fields have equal magnitudes, i.e. $|\vec E|=|\vec B|$%
, so that particle pair production due to a strong electric field near the
string is suppressed. A basic ansatz which describes a straight
superconducting string has been examined. For this basic ansatz, the modulus
of the scalar condensate field, $|\phi |$, and the gauge field structure
function $P$, are considered to be functions only of the radial distance $r$
from the center of the string core, and the phase $\psi $ of the scalar
condensate depends only upon the coordinate $z$ along the string and the
time $t$. Fluctuations of the scalar and vector fields about the primary
fields of the basic ansatz have been considered, and linear field equations
for these fluctuation, or excitation, fields have been obtained.

The behavior of the excitation fields is determined by a set of nontrivial
coupled equations, and approximate solutions have been obtained by making
certain approximations and imposing simplifying conditions, such as $%
Q^2-I^2=0$ ($\partial _\mu \psi \partial ^\mu \psi =0$), along with others.
These solutions are {\it particular} solutions, which can be associated with
possible {\it nonradiative} excitation modes for the scalar and vector
fields of the model, and are expected to be special cases of more general
solutions, which may include descriptions of {\it radiative} excitation
modes. The equation of motion for the scalar excitation field $H$ appears to
describe a massive `Higgs' mode, but, because of an auxiliary condition
which has been imposed, the particular solutions obtained describe an
effectively massless scalar mode which propagates along the string with the
primary current pulse at the speed of light. Similarly, the equation of
motion for the vector field $a^\mu $ appears to describe a massive photon,
but the particular solutions obtained describe an effectively massless
vector field which propagates along the string with the scalar field. In
each case, the effect of the mass parameter is to modulate the radial
profile of the effectively massless mode. The nonvanishing solutions for $%
H_l $ and $a_l^\mu $ have nonzero angular momentum contributions, i.e. $%
l\neq 0,\,1$.

Both the scalar and photon excitation fields contribute to the current
density $j^\mu $ and thus serve as sources for electromagnetic fields. The
current density $j^\mu $ generates an electric field with components $e_r$
and $e_\theta $, and a magnetic field with components $b_r$ and $b_\theta $.
Furthermore, since $j_l^\mu $ vanishes for $l=0$, the total charge per unit
length $q$ and the total current $i$ generated by $j^\mu $ vanish, with $%
q=i=0$. The Poynting vector $\vec s$ resulting from $\vec e$ and $\vec b$
possesses a nonvanishing component only in a direction parallel to the
string, so that $s_r=0$, indicating that the string does not absorb or emit
electromagnetic radiation. The particular solutions obtained therefore
describe nonradiative excitation modes.

It can be noted that $H(\xi )$ and $a^\mu (\xi )$ are somewhat arbitrary
functions of $\xi $ (subject to the requirement that they be sufficiently
small bounded functions of $\xi $), with their profiles along the string
presumably being dictated by initial conditions, which, in turn, are
produced by interactions of the superconducting string with other strings,
charged particles, or electromagnetic fields. Therefore, if the
superconducting string experiences any electromagnetic interactions,
excitation fields are expected to be produced. It is hoped that the special
class of excitation field solutions studied here can yield some insight
regarding more general phenomena involving superconducting cosmic strings.

\end{document}